\documentclass{IEEEtran4PSCC}

\usepackage{stfloats}
\usepackage{url}
\usepackage{verbatim}
\usepackage{graphicx}
\usepackage{cite}
\usepackage[utf8]{inputenc}

\usepackage{hyperref}
\hypersetup{hidelinks}
\usepackage{xcolor,soul,framed} 

\usepackage{amsfonts,amssymb}
\usepackage{mathrsfs}
\usepackage{amsthm}
\usepackage[cmex10]{amsmath}
\usepackage{mathtools}
\usepackage{siunitx}
\usepackage{bbm}
\DeclareMathAlphabet{\mathbbb}{U}{bbold}{m}{n}
\usepackage{array}
\usepackage{physics}
\usepackage{cite}
\usepackage{flushend}
\usepackage{nomencl}
\usepackage{makecell}
\usepackage{threeparttable}
\usepackage{colortbl}

\usepackage{enumitem}

\makenomenclature

\theoremstyle{remark}

\newtheorem{example}{Example}

\newcommand{\trans}     {^{\mathsf{T}}}
\newcommand{\pha}[1]    {\underline{#1}}
\newcommand{\phaconj}[1]{\overline{\underline{#1}}}

\newcommand{\phavec}[1] {\underline{\boldsymbol{#1}}}

\newcommand{\mat}[1]    {\boldsymbol{#1}}
\newcommand{\phamat}[1] {\underline{\boldsymbol{#1}}}

\makeatletter
\let\old@ps@headings\ps@headings
\let\old@ps@IEEEtitlepagestyle\ps@IEEEtitlepagestyle
\def\psccfooter#1{%
    \def\ps@headings{%
        \old@ps@headings%
        \def\@oddfoot{\strut\hfill#1\hfill\strut}%
        \def\@evenfoot{\strut\hfill#1\hfill\strut}%
    }%
    \def\ps@IEEEtitlepagestyle{%
        \old@ps@IEEEtitlepagestyle%
        \def\@oddfoot{\strut\hfill#1\hfill\strut}%
        \def\@evenfoot{\strut\hfill#1\hfill\strut}%
    }%
    \ps@headings%
}
\makeatother

\psccfooter{%
        \parbox{\textwidth}{\hrulefill \\ \small{23rd Power Systems Computation Conference} \hfill \begin{minipage}{0.2\textwidth}\centering \vspace*{4pt} \includegraphics[scale=0.06]{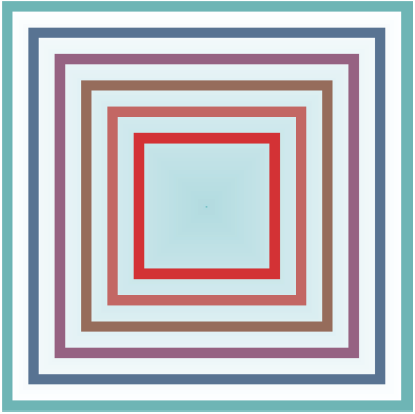}\\\small{PSCC 2024} \end{minipage} \hfill \small{Paris, France --- June 4 -- 7, 2024}}%
}

\begin{document}
%
\title{Saturation-Informed Current-Limiting Control \\  for Grid-Forming Converters}

\author{
\IEEEauthorblockN{Maitraya Avadhut Desai, Xiuqiang He, Linbin Huang, Florian Dörfler\\}
\IEEEauthorblockA{Automatic Control Laboratory, ETH Zurich, 8092 Zurich, Switzerland \\
mdesai@student.ethz.ch, xiuqhe@ethz.ch, linhuang@ethz.ch, dorfler@ethz.ch \\}
}


\maketitle

\begin{abstract}

In this paper, we investigate the transient stability of a state-of-the-art grid-forming complex-droop control (i.e., dispatchable virtual oscillator control, dVOC) under current saturation. We quantify the saturation level of a converter by introducing the concept of degree of saturation (DoS), and we propose a provably stable current-limiting control with saturation-informed feedback, which feeds the degree of saturation back to the inner voltage-control loop and the outer grid-forming loop. As a result, although the output current is saturated, the voltage phase angle can still be generated from an internal virtual voltage-source node that is governed by an equivalent complex-droop control. We prove that the proposed control achieves transient stability during current saturation under grid faults. We also provide parametric stability conditions for multi-converter systems under grid-connected and islanded scenarios. The stability performance of the current-limiting control is validated with various case studies.

\end{abstract}


\section{Introduction}\label{sec:Intro}
Grid-forming (GFM) control is seen as a promising solution to smoothly transition from synchronous machine-dominated power systems to converter-interfaced renewable energy sources-dominated power systems \cite{Milano-review}. GFM converters are controlled as voltage sources behind impedance during stable operation \cite{lasseter2019grid}, thus contributing to the strength of the grid by autonomously regulating voltage and frequency.

Numerous GFM control strategies have been proposed in the literature. Power-frequency (p-f) droop control is the basis of all the earliest types of GFM controls \cite{Chandorkar93}. Another typical variant is virtual synchronous machine (VSM) \cite{Beck2007VirtualSM}, which seeks to mimic synchronous machines. Although these well-known GFM control strategies render converters adaptable to the legacy power system, they follow an elementary single-input single-output (SISO) design philosophy. In recent years, the dispatchable virtual oscillator control (dVOC) GFM strategy has been developed using a multivariate design philosophy in \cite{Colombino-dVOC,Gross-dVOC,Subotic-dVOC} and \cite{Lu_dVOC} by top-down and bottom-up design procedures, respectively. The dVOC effectively handles the coupling between two interactive loops (i.e., p-f and Var-Volt) with a multivariate control architecture. This control strategy works beyond the conventional SISO design of the p-f droop control or VSM. Therefore, it is considered a promising variant of GFM controls \cite{Lu-benchmarking}. The dVOC has been shown to be an extension of the classical p-f droop control from the perspective of a recently developed concept termed {complex frequency} \cite{Milano-complex-freq}, thus also referred to as \textit{complex-droop control} \cite{He-cplx-freq-sync}. The stability of a multi-converter system with complex-droop control (i.e., dVOC) has been analytically investigated, and parametric stability conditions are provided in \cite{He-cplx-freq-sync, Xiuqiang_NonlinearStab_dVOC, Xiuqiang_Preprint, he2024passivity}.

Although a GFM converter behaves as a voltage source behind an impedance, its current output needs to be limited under large disturbances to protect the converter, which is very different from conventional synchronous machines. To be specific, when subjected to grid disturbances, e.g., grid faults, synchronous machines can provide up to 5-7~p.u. fault current. However, semiconductor-based converters can only provide a fault current of about 1.2~p.u. \cite{MicrogridLasseter}. Hence, to protect converters from overcurrent damage, various current-limiting strategies have been developed in the literature, typically including: 1) threshold/adaptive virtual impedance \cite{paquette2015virtual,qoria2020current}, 2) current limiter with virtual admittance \cite{rosso2021implementation,fan2022equivalent,zhang2023current,saffar2023impacts,laba2023virtual}, and 3) current-forming voltage-following control \cite{li2022revisiting,SEPFCLinbin}. 

Under grid faults, it is essential that the converters maintain transient stability for a successful fault ride-through (FRT) \cite{Wu-PSC}. Therefore, the transient stability of GFM converters, considering the aforementioned current-limiting strategies, is of great significance. The effect of a prioritized current limiter on the transient stability of p-f droop-controlled converters has been studied in \cite{SEPFCLinbin}, and an enhancement to the GFM control was proposed to render the converter stable during current saturation. The transient stability of p-f droop-based converters with a circular current limiter was studied in \cite{fan2022equivalent}. To enhance the transient stability of p-f droop-based converters with a circular current limiter, \cite{laba2023virtual} proposed the utilization of the unsaturated power measurement for the active power feedback. In \cite{gross2023towards,samanta2023nonlinear}, current-limiting constraints were incorporated into an optimal control problem to address overcurrent. Moreover, the transient stability of VSMs under current saturation was investigated in \cite{VSMCurrLim_Darbandi, VSG_trans_currlim_Pingjuan}. However, all the aforementioned studies consider transient stability using SISO loops, i.e., in terms of p-f mapping. Recently, efforts have also been made to study multivariate GFM controlled converters considering current limitation. In \cite{ModelRednAjala2021}, a reduced-order model of dVOC-based converters with circular current limiters was developed. A phase-plane analysis for the transient stability examination of dVOC-based converters considering current limitation was presented in \cite{MAwal21}. Hybrid angle control (HAC), which is similar to dVOC, was developed with a novel current-limiting strategy in \cite{HybridAngle_Tayyebi}. However, the transient stability of dVOC-based GFM converters under current limitation remains underexplored due to the intricate nature of the multivariate control architecture and the presence of saturation nonlinearity. Moreover, it is unclear how to design a stable current-limiting control under such a context, especially when a \textit{multi-converter} system is considered.

To this end, this paper aims to analyze the transient stability of GFM converters with complex-droop control (i.e., dVOC), taking into consideration the circular current-limiting strategy. Furthermore, we focus on developing a stability-guaranteed current-limiting control for GFM converters under current saturation. This paper has three substantial contributions. First, an equivalent circuit of the dVOC-based GFM converter under current saturation is developed for control design and stability analysis. Second, a saturation-informed current-limiting control is proposed to ensure stability under current saturation of GFM converters. Third, parametric conditions are obtained to ensure the transient stability of the proposed control for a system that contains one or more GFM converters.

\section{System Description and Equivalent Circuit}\label{sec:Complex_Droop_Control}

\subsection{Configuration of Complex-Droop Control}

\begin{figure}
    \centering
    \includegraphics{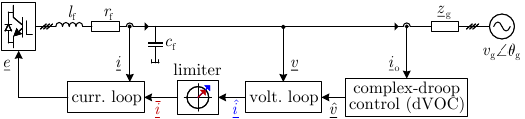}
     \caption{A grid-forming converter with complex-droop control (i.e., dVOC) and a circular current limiter.}
    \label{fig:ComplexDroopSimple}
\end{figure}

The model of a grid-connected GFM converter is shown in Fig.~\ref{fig:ComplexDroopSimple}, where the complex-droop control (i.e., dVOC) is employed. The inner controllers track the reference provided by the complex-droop control. A circular current limiter is placed between the inner controllers to modify the current reference to achieve the current limitation. The three-phase system is assumed to be balanced, hence, enabling the use of $\alpha\beta$ or $dq$ coordinates based on the Clarke or Park transformations, respectively. For the modeling and analysis in this paper, the complex-vector coordinates, e.g., $\pha v \coloneqq v_{\alpha} + j v_{\beta}$, are used.

\subsubsection{Complex-Droop Control as GFM Reference Model}
The reference voltage signal is provided by the complex-droop control (i.e., dVOC) as \cite{Colombino-dVOC}
\begin{equation}
\label{eq:Columbino-dVOC}
    {\pha{\dot{\hat{v}}}} = j{\omega}_0 {\pha{\hat{v}}} + \eta {e^{j\varphi}}\Bigl( \frac{p^{\star} - jq^{\star}}{{{v}^{\star 2}}} {\pha{\hat{v}}} - {\pha{i}}_{\rm o} \Bigr) +  \eta \alpha\Bigl( \frac{{{v}^{\star 2} - \abs{\pha{\hat{v}}}^2}}{{{v}^{\star 2}}}\Bigr){\pha{\hat{v}}},
\end{equation}
where the underline represents complex-vector variables, $\omega_0$ denotes the fundamental frequency, ${\pha{i}}_{\rm o}$ denotes the measured output current, $p^{\star}$, $q^{\star}$, and $v^{\star}$ denote the active power, reactive power, and voltage amplitude setpoints, respectively. The exponential term $e^{j\varphi}$ with $\varphi \in [0,\pi/2]$ denotes the rotation operator to adapt to the network impedance characteristics. Finally, $\eta,\alpha \in \mathbb{R}_{\geq 0}$ are control gains. In \eqref{eq:Columbino-dVOC}, the control law can be understood as follows: The first term induces harmonic oscillations of frequency $\omega_0$, the second term synchronizes the relative phase angles to the power setpoints via the measured current feedback, and the third term regulates the voltage amplitude using a quadratic function. As the complex-droop control subsumes a power-frequency droop relationship \cite{He-cplx-freq-sync}, it is backward compatible with the operation of pre-existing GFM controllers and legacy power systems.

\subsubsection{Inner Controllers and Current Limiter}
The inner controllers can either be implemented in the $\alpha\beta$ coordinate frame with proportional resonant (PR) control or in the $dq$ coordinate frame with proportional integral (PI) control \cite{Subotic-dVOC}. These two implementations are approximately equivalent around the nominal frequency, i.e., $\dot{\theta} \approx \omega_0$. The results of this work apply to both types of control implementations.
 
The inner voltage controller consists of a proportional and resonant (PR) regulator and feedforward components as
\begin{subequations}
\label{VoltageControl}
\begin{align}
    {\pha{\dot{\zeta}}_{\rm v}} &= j \omega_0 {\pha{{\zeta}}_{\rm v}} + {\pha{\hat{v}}}- {\pha{v}},\label{eq:IV}\\
    \pha{\hat i} &= k_{\rm p}^{\rm v} \bigl({\pha{\hat{v}}}- {\pha{v}}\bigr) + k_{\rm r}^{\rm v}\pha{\zeta}_{\rm v} + \pha{i}_{\rm o} + j\omega_0 c_{\rm f} \pha{v},
    \label{PIopv}
\end{align}
\end{subequations}
where $k_{\rm p}^{\rm v},\, k_{\rm r}^{\rm v} \in \mathbb{R}_{> 0}$ are the control gains, $\pha{\zeta}_{\rm v}$ is the integrator state, $\pha{v}$ is the voltage across the filter capacitance $c_{\rm f}$, and the terms $ \pha{i}_{\rm o} + j\omega_0 c_{\rm f}\pha{v}$ represent the feedforward components, which are optional to use. Alternatively, the voltage controller can also use a complex-gain proportional regulator to generate the current reference, which acts as a virtual admittance, particularly in the case of current saturation. Specifically, the current reference is given as \cite{rosso2021implementation,fan2022equivalent,zhang2023current,saffar2023impacts,laba2023virtual},
\begin{equation}
\label{VirtualAdmittance}
    \pha{\hat i} = \frac{1}{\pha{z}_{\rm v}}\left({\pha{\hat{v}}}- {\pha{v}}\right),
\end{equation}
where $\pha{z}_{\rm v}$ serves as the virtual impedance behind the terminal.

The circular current limiter is used to protect the converter from overcurrent damage. The current limiter scales down the current reference magnitude only when the magnitude of $\pha{\hat i}$ is larger than a certain threshold $i_{\lim}$ while keeping its phase angle unchanged, and the output is denoted as $\pha{\bar i}$, given by
\begin{gather}
    \label{eq:CircCurrLim}
    \pha{\bar i} =
    \begin{cases}
        \pha{\hat i} & \text{if } \lvert\pha{\hat i}\rvert \leq i_{\lim},\\
          \frac{i_{\lim}}{\lvert\pha{\hat i}\rvert}\pha{\hat i} & \text{if } \lvert\pha{\hat i}\rvert > i_{\lim}.
    \end{cases}
\end{gather}

The inner current controller consists of a proportional and resonant (PR) regulator and feedforward components as
\begin{subequations}
\begin{align}
 {\pha{\dot{\zeta}}_{\rm c}} &= j \omega_0 {\pha{{\zeta}}_{\rm c}} + {\pha{\bar i}}- {\pha{i}},\label{eq:IC}\\
    \pha{e} &= k_{\rm p}^{\rm c}\bigl({\pha{\bar i}}- {\pha{i}}\bigr) + k_{\rm r}^{\rm c}\pha{\zeta}_{\rm c} + \pha{v} + j\omega_0 l_{\rm f}\pha{i} + r_{\rm f}\pha{i},
    \label{PIopc}
\end{align}
\end{subequations}
where $k_{\rm p}^{\rm c},\, k_{\rm r}^{\rm c} \in \mathbb{R}_{> 0}$ are the control gains, $\pha{\zeta}_{\rm c}$ is the integrator state, $ \pha{i}$ is the current passing through the filter inductance $l_{\rm f}$ and filter resistance $r_{\rm f}$, and the terms $\pha{v} + j\omega_0 l_{\rm f}\pha{i} + r_{\rm f}\pha{i}$ represent the feedforward components (optional to use).

\subsection{Equivalent Circuit Under Current Saturation}

For ease of control design and transient stability analysis, we ignore the electromagnetic dynamics in the circuit, as well as the dynamics of the inner current control loop, which leads to ${\pha{i}}={\pha{\bar i}}$. These assumptions are standard and commonly made for transient stability analysis, as in \cite{SEPFCLinbin}. The primary concern in our study is the dynamics of the GFM control loop and the voltage control loop.

When the current limiter is not triggered, the inner controllers track the reference signal provided by the reference model, i.e., ${\pha{\hat{v}}}={\pha{v}}$, or a virtual impedance is emulated, i.e., $\pha{\hat{v}} = \pha{v} + \pha{z}_{\rm v}\pha{i}$. A complete transient stability analysis in this case can be found in \cite{Xiuqiang_Preprint}. However, when the current limiter is triggered due to disturbances or faults in the grid, the reference cannot be effectively tracked by the inner controllers. To avoid the windup of the PR voltage controller, the integrator can be disabled directly. The feedforward components can also be turned off to simplify analysis and control design. Thus, during current saturation, the voltage control in \eqref{VoltageControl} becomes a particular case of the virtual admittance control in \eqref{VirtualAdmittance}, with $\pha{z}_{\rm v} = 1/k_{\rm p}^{\rm v} \in \mathbb{R}_{>0}$ acting as virtual resistance.
\begin{figure}
    \centering
    \includegraphics{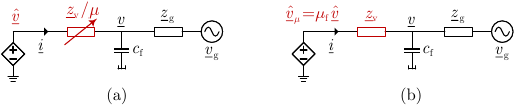}
     \caption{(a) Current saturation leads to current-dependent virtual impedance in the equivalent circuit \cite{fan2022equivalent}. (b) The proposed current-limiting control leads to a constant virtual impedance and a virtual voltage-source node with current-dependent adaptive magnitude under current saturation.}
    \label{fig:EqCktdVOCSatCirc}
\end{figure}

We define the \textit{degree of saturation (DoS)} of the current limiter in \eqref{eq:CircCurrLim} with the ratio between the output and the input,
\begin{equation}
    \mu \coloneqq \frac{\, \pha{\bar i}\, }{\pha{\hat i}} \in (0,\, 1].
    \label{DoSDef}
\end{equation}
When the current is unsaturated, $\pha{\bar i} = \pha{\hat i}$, i.e., $\mu = 1$. When saturated, however, $0 < \mu < 1$ since $\lvert\pha{\hat i}\rvert > i_{\lim} = \lvert \pha{\bar i} \rvert $. From \eqref{VirtualAdmittance}, \eqref{DoSDef} and the assumption $\pha{i} = {\pha{\bar i}}$, it follows that 
\begin{equation}
    \pha{i} = \pha{\bar i} = \mu \pha{\hat i} = \mu \frac{1}{\pha{z}_{\rm v}}\left({\pha{\hat{v}}}- {\pha{v}}\right),
    \label{PIopcurropsat}
\end{equation}
which establishes a relationship between the converter output current, the terminal capacitor voltage, and the reference voltage. Based on~\eqref{PIopcurropsat}, we obtain an equivalent circuit of the GFM converter under current saturation, as shown in Fig.~\ref{fig:EqCktdVOCSatCirc}(a). The equivalent virtual impedance is $\pha{z}_{\rm v}/\mu$, with $\mu$ depending on the current reference. Since the equivalent impedance is current-dependent and thus time-varying during transients, it is nontrivial to develop an analytical stability condition for the converter under current saturation. Moreover, during severe saturation scenarios, the DoS may drop to a very low value, resulting in a very high virtual impedance. This is detrimental to the transient stability of the converter \cite{Xiuqiang_Preprint}. We aim to overcome this limitation by introducing additional feedback loops to obtain a constant impedance in the equivalent circuit.

\section{Saturation-Informed Complex-Droop Control}\label{sec:Sat-Inform_Control}

\subsection{Saturation-Informed Feedback Control Design}

\begin{figure}
    \centering
    \includegraphics[width=\linewidth]{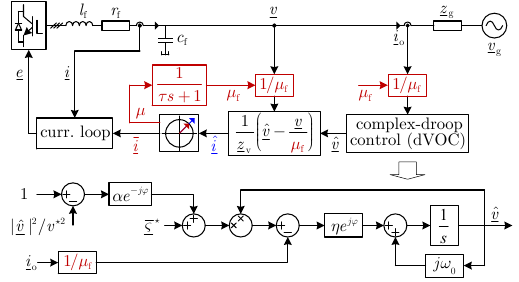}
     \caption{The proposed saturation-informed current-limiting control strategy scales up the current and voltage feedback signal by the DoS $\mu_{\rm f}$.}
    \label{fig:SatInf-GFM}
\end{figure}

During a voltage dip and current saturation, the converter terminal voltage ``drops down" while the output current is limited, preventing the GFM control from operating normally. To ensure that the GFM control works as if the current were not saturated, we feed the DoS signal back to the GFM control loop \emph{and} the inner voltage control loop to ``scale up" the feedback signals, as demonstrated by the gain $1/\mu_{\rm f}$ in Fig.~\ref{fig:SatInf-GFM}.

A low-pass filter is used to eliminate the algebraic loop in the proportional voltage control loop. The filter also aims to make the voltage control slower than the current control. It is represented as
\begin{equation}
        \label{eq:filtDoS}
        {\mu}_{\rm f} = \frac{1}{\tau s + 1} \mu,
\end{equation}
where $\tau \in \mathbb{R}_{>0}$ is the filter time constant.

Since the current feedback is scaled by the DoS signal, the complex-droop control law \eqref{eq:Columbino-dVOC} is modified using $\mu_{\rm f}$ as
\begin{equation}
    \label{eq:currfbdVOCSat}
    {\pha{\dot{\hat{v}}}} = j{\omega}_0 {\pha{\hat{v}}} + \eta {e^{j\varphi}}\Bigl( \frac{p^{\star} - jq^{\star}}{{{v}^{\star 2}}} {\pha{\hat{v}}} - \frac{{\pha{i}}_{\rm o}}{\mu_{\rm f}} \Bigr) + \eta \alpha \Bigl(\frac{{{v}^{\star 2} - \abs{\pha{\hat{v}}}^2}}{{{v}^{\star 2}}}\Bigr){\pha{\hat{v}}}.
\end{equation}
The feedback voltage in the inner voltage controller is also scaled by the DoS. Thus, \eqref{VirtualAdmittance} is modified as

\begin{equation}
        \pha{\hat i} = \frac{1}{\pha{z}_{\rm v}}\Bigl({\pha{\hat{v}}}- \frac{{\pha{v}}}{\mu_{\rm f}}\Bigr),
        \label{PIopvsatMod_stable}
\end{equation}
with saturation-informed feedback $1/\mu_{\rm f}$. As seen in \eqref{eq:currfbdVOCSat} and \eqref{PIopvsatMod_stable}, the additional feedback introduced by $1/\mu_{\rm f}$ can still preserve negative feedback characteristics during current saturation. To enable the current-limiting strategy, a switching of the voltage controller from the form \eqref{VoltageControl} or \eqref{VirtualAdmittance} to the form \eqref{PIopvsatMod_stable} is necessary during current saturation and from \eqref{PIopvsatMod_stable} to \eqref{VoltageControl} or \eqref{VirtualAdmittance} after recovery. For the purpose of the switching, the voltage magnitude $\lvert\pha{v}\rvert$ can be compared with a predetermined limit $v_{\rm sat}$ to start an FRT process. The threshold $v_{\rm sat}$ may depend on the specification of grid codes, e.g., 0.9 p.u. Alternatively, one may also use the DoS $\mu_{\rm f}$ and/or the combination of multiple signals for this switch.

To examine how the modified controllers work, we further define an internal virtual voltage as 
\begin{equation}
\label{eq:defVirtualVoltage}
    \hat{\pha v}_{\mu} \coloneqq \mu_{\rm f} \hat{\pha v}.
\end{equation}
From \eqref{DoSDef}, we obtain $\pha{i} = \pha{\bar i} = \mu \pha{\hat i} \approx \mu_{\rm f} \pha{\hat i}$, while ignoring the dynamics of the low-pass filter. Considering \eqref{PIopvsatMod_stable} and \eqref{eq:defVirtualVoltage}, the converter output current can therefore be written as
\begin{equation}
    \pha{i} =  \frac{1}{\pha{z}_{\rm v}}\left({\pha{\hat{v}}_{\mu}}-{\pha{v}}\right),
    \label{eq:PIopcurropsat_stable}
\end{equation}
which leads to a new equivalent circuit, as shown in Fig.~\ref{fig:EqCktdVOCSatCirc}(b). Using \eqref{eq:currfbdVOCSat} and \eqref{eq:defVirtualVoltage}, an equivalent complex-droop control that governs the internal virtual voltage is obtained as
\begin{equation}
    \label{eq:AuxDVOCModEquiv}
    {\pha{\dot{\hat{v}}}_{\mu}} = j{\omega}_0 {\pha{\hat{v}}_{\mu}} + \eta {e^{j\varphi}}\Bigl( \frac{p^{\star} - jq^{\star}}{{{v}^{\star 2}}} {\pha{\hat{v}}_{\mu}} - {{\pha{i}}_{\rm o}} \Bigr) + \eta \alpha \Bigl(\frac{{{v}_{\mu}^{\star 2} - \abs{\pha{\hat{v}}_{\mu}}^2}}{{v_{\mu}^{\star 2}}}\Bigr){\pha{\hat{v}}_{\mu}},
\end{equation}
where we ignore the term $\dot{\mu_{\rm f}}\hat{\pha v}$ as the dynamics of the filtered DoS is ignored for the ease of analysis. The modified setpoint is denoted as $v_{\mu}^{\star} \coloneqq \mu_{\rm f} {v}^{\star}$. This equivalent ``controller" in \eqref{eq:AuxDVOCModEquiv} regulates the internal virtual voltage $\pha{\hat{v}}_{\mu}$ in Fig.~\ref{fig:EqCktdVOCSatCirc}(b). 

As a consequence of using \eqref{eq:currfbdVOCSat} and \eqref{PIopvsatMod_stable}, a constant virtual impedance is obtained in the equivalent circuit of the GFM converter under current saturation, as shown in Fig.~\ref{fig:EqCktdVOCSatCirc}(b). Moreover, the internal virtual voltage source is adaptively modulated by the DoS. Although the output current is saturated, the phase angle of the terminal voltage is still controlled by the internal virtual voltage-source node. This is because \eqref{eq:defVirtualVoltage} ensures the same angle for the internal virtual voltage and the reference voltage. In other words, the proposed saturation-informed control maintains the synchronizing and stabilizing capability while limiting the output current. While the saturation-informed feedback control is developed considering complex-droop control, this method also applies to other types of GFM control such as classical droop control and VSM.

\subsection{Existence of Current-Saturated Steady-State Equilibrium}
We ignore the large impedance of the filter capacitor and denote the equivalent series impedance of the grid-connected GFM converter as $\pha{z} \coloneqq \pha{z}_{\rm g} + {\pha{z}_{\rm v}}$. The equivalent admittance is denoted as $\pha{y} = {1}/{\pha{z}}$. Furthermore, we denote the rotated grid-impedance angle by $\phi \coloneqq \angle{\pha{z}} - \varphi$. Both $\lvert \pha{y} \rvert$ and $\phi$ are constant. To derive the steady-state equations, we further define the rotated power setpoints, the phase angle with respect to the grid, and the grid frequency deviation by $\rho_{\varphi}^{\star} + j \sigma_{\varphi}^{\star} \coloneqq e^{j\varphi} \frac{p^{\star} - jq^{\star}}{v^{\star 2}}$, $\delta \coloneqq \angle {\pha {\hat v}}_{\mu} - \theta_{\rm g}$, and $\omega_{\Delta} \coloneqq \omega_0 - \omega_{\rm g}$ respectively. We denote $\hat v_{{\mu}\rm s}$, $\delta_{\rm s}$ and $\mu_{\rm s}$ as the steady state of $\lvert \pha{\hat{v}}_{\mu} \rvert$, $\delta$, and $\mu_{\rm f}$, respectively.

By performing similar manipulations as in \cite[Lemma~3]{Xiuqiang_Preprint}  and using the cosine theorem of the circuit voltage-vector triangle, we can obtain the set of current-saturated steady-state equations for the grid-connected saturation-informed complex-droop-controlled converter as
\begin{align}
\label{eq:saturation-feedback-control}
    \begin{aligned}
        \sigma_{\varphi}^{\star} + \alpha - \alpha \tfrac{\hat v_{{\mu}\rm s}^2}{\mu_{\rm s}^2 {v}^{\star 2}} & =  \lvert \pha{y} \rvert\cos{\phi} - v_{\rm g} \lvert \pha{y} \rvert \cos \left(\delta_{\rm s} + \phi \right) /\hat v_{{\mu}\rm s}, \\
        \rho_{\varphi}^{\star} + \tfrac{\omega_{\Delta}}{\eta} & = - \lvert \pha{y} \rvert\sin{\phi} + v_{\rm g} \lvert \pha{y} \rvert \sin \left(\delta_{\rm s} + \phi \right) /\hat v_{{\mu}\rm s}, \\
        \cos \delta_{\rm s} &= \tfrac{\hat v_{{\mu}\rm s}^2 + v_{\rm g}^2 - i_{\lim}^2 \lvert \pha{z} \rvert ^2}{2 \hat v_{{\mu}\rm s} v_{\rm g}}.
    \end{aligned}
\end{align}
The explicit conditions for the existence of a solution of \eqref{eq:saturation-feedback-control} can be analytically derived. Briefly, the last two equations can be used to eliminate $\delta_{\rm s}$ and obtain the conditions for the existence of $\hat v_{{\mu}\rm s}$. Consequently, the following relationship
\begin{equation}
    (\sigma_{\varphi}^{\star} + \alpha - \alpha \tfrac{\hat v_{{\mu}\rm s}^2}{\mu_{\rm s}^2 {v}^{\star 2}})^2 + (\rho_{\varphi}^{\star} + \tfrac{\omega_{\Delta}}{\eta})^2 = \tfrac{i_{\lim}^2}{\hat v_{{\mu}\rm s}^2},
\end{equation}
which follows from \eqref{eq:saturation-feedback-control}, can be used to derive the conditions for the existence of $\mu_{\rm s}$. We omit the lengthy manipulation for the general case. In what follows, we illustrate as an example, that the existence of an equilibrium can be easily ensured under certain parametric conditions. This further provides practical guidelines for parameter tuning of saturation-informed complex-droop-controlled converters.

\begin{example}
\label{specialCase}
Consider the following parametric tuning for the saturation-informed complex-droop-controlled converter during current saturation: $\varphi = \angle {\pha{z}_{\rm v}} = \angle {\pha z_{\rm g}}$ (the same impedance angle), $\rho_{\varphi}^{\star} = 0$ (reactive power receives the highest priority), $\omega_{\Delta} = 0$ (nominal grid frequency).
\end{example}

For Example~\ref{specialCase}, the set of steady-state equations in \eqref{eq:saturation-feedback-control} always has a proper solution, where $\mu_{\rm s} \in \mathbb{R}_{>0}$, if $\sigma_{\varphi}^{\star} + \alpha > \frac{i_{\lim}}{v_{\rm g} + i_{\lim}\lvert \pha {z} \rvert}$, or $(\sigma_{\varphi}^{\star} + \alpha) (\lvert \pha {z}_{\rm g} \rvert + \lvert \pha{z}_{\rm v} \rvert) > 1$ (without relying on grid voltage information) or simply $(\sigma_{\varphi}^{\star} + \alpha) \lvert \pha{z}_{\rm v} \rvert \geq 1$ (without relying on grid voltage and impedance information). This statement can be easily justified. Under the above parametric conditions, we observe $\phi = 0$, $\delta_{\rm s} = 0$, and $\hat v_{{\mu}\rm s} = v_{\rm g} + i_{\lim} \lvert \pha {z} \rvert$. Therefore, the first equation in \eqref{eq:saturation-feedback-control} simplifies to $   \sigma_{\varphi}^{\star} + \alpha - \alpha \tfrac{\hat v_{{\mu}\rm s}^2}{\mu_{\rm s}^2 {v}^{\star 2}} =  \tfrac{i_{\lim}}{v_{\rm g} + i_{\lim}\lvert \pha {z} \rvert}$.
The existence of $\mu_{\rm s}$ requires that $\frac{\hat v_{{\mu}\rm s}^2}{\mu_{\rm s}^2 {v}^{\star 2}} > 0$, i.e.,
$    \sigma_{\varphi}^{\star} + \alpha > \tfrac{i_{\lim}}{v_{\rm g} + i_{\lim}\lvert \pha {z} \rvert}$.
Since $\pha{z} = \pha{z}_{\rm g} + \pha{z}_{\rm v}$, where $\angle {\pha z_{\rm g}} = \angle {\pha{z}_{\rm v}}$, it follows that
$   \frac{1}{ \lvert \pha{z}_{\rm v} \rvert} > \tfrac{1}{\lvert \pha{z}_{\rm g} \rvert + \lvert \pha{z}_{\rm v} \rvert} = \tfrac{1}{ \lvert \pha {z} \rvert } \geq \tfrac{i_{\lim}}{v_{\rm g} + i_{\lim}\lvert \pha {z} \rvert}$.
Therefore, both the parametric conditions $(\sigma_{\varphi}^{\star} + \alpha) (\lvert \pha {z}_{\rm g} \rvert + \lvert \pha{z}_{\rm v} \rvert) > 1$ and $ (\sigma_{\varphi}^{\star} + \alpha) \lvert \pha{z}_{\rm v} \rvert \geq 1 $ can guarantee the existence of a solution where $\mu_{\rm s} \in \mathbb{R}_{>0}$.

We note that if the exact information of the grid-impedance $\pha{z}_{\rm g}$ is available, e.g. through grid-impedance estimation methods \cite{Fang_GridImpedanceEst_2021},  \cite{Wang_GridImpedanceDetection_2023}, it can be used in the above analysis and guarantee. If the information is not available, we may consider the grid impedance in the worst case or use a sufficient condition without relying on the grid impedance information, e.g., the last condition discussed above.

\subsection{Relationship Between Original and Augmented Networks} 
\begin{figure}
    \centering
    \includegraphics{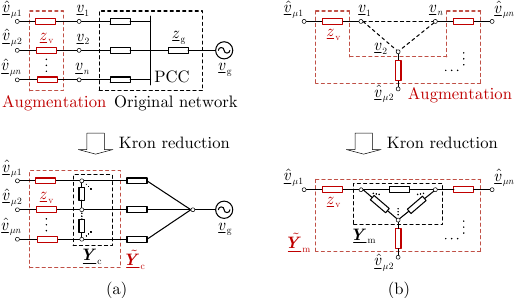}
     \caption{Kron reduction of the original network for (a) multi-converter grid-connected systems and (b) multi-converter islanded microgrids.}
    \label{fig:Multi_MuG_Admittance}
\end{figure}

The use of the saturation-informed control strategy in multi-converter grid-connected systems or multi-converter islanded microgrids, during current saturation, leads to an augmentation of the original network with the additional virtual impedance $\pha{z}_{\rm v}$ (similar to the case of single-converter grid-connected systems). The augmented network considers the constant virtual impedance $\pha{z}_{\rm v}$, as shown in Fig.~\ref{fig:Multi_MuG_Admittance}.

We consider the case of multi-converter grid-connected systems as an example to formulate the relationship between the original network and the augmented network. The output current $\phavec i_{\rm o}$, a vector corresponding to the output of multiple converters (similar notation is used below), can be expressed in terms of the chosen virtual voltage $\phavec {\hat{v}}_{\mu}$  and the grid voltage $\pha v_{\rm g}$ with the collector network equation as
\begin{equation}
\label{eq:collector}
    \begin{bmatrix}
    \phavec{i}_{{\rm o}} \\ \pha{i}_{\rm g}
    \end{bmatrix} =
    \phamat{\Tilde{Y}}_{\rm red}
    \begin{bmatrix}
        \phavec{\hat{v}}_{\mu} \\ \pha{v}_{\rm g}
    \end{bmatrix}, \quad \phamat{\Tilde{Y}}_{\rm red} = \begin{bmatrix}
        \phamat{\Tilde{Y}}_{\rm c} & -\phavec{\Tilde{y}} \\
        -\phavec{\Tilde{y}}\trans & \pha{\Tilde{y}}_{\rm g}
    \end{bmatrix},
\end{equation}
where $\phamat{\Tilde{Y}}_{\rm red} \in \mathbb{C}^{(n+1) \times (n+1)}$ is the admittance matrix of the Kron-reduced augmented network whereas $\phamat{\Tilde{Y}}_{\rm c} \in \mathbb{C}^{n \times n}$ is known as the nonsingular admittance matrix (in which the grid node and the associated elements $\phavec{\Tilde{y}}$ and $\pha{\Tilde{y}}_{\rm g}$ are excluded). For the original network without including $\pha z_{\rm v}$, the original nonsingular admittance matrix is denoted as $\phamat{Y}_{\rm c}$.
Analogous to \eqref{eq:collector}, the output current can be written in terms of the terminal voltage and the grid voltage as
$    \phavec i_{\rm o} = \phamat{Y}_{\rm c} \phavec v - \phavec{y} \pha v_{\rm g}$.
Furthermore, the terminal voltage $\phavec v$ and the internal virtual voltage $\phavec{\hat{v}}_{\mu}$ are related as
$    \phavec v = \phavec{\hat{v}}_{\mu} - \phavec i_{\rm o}{\pha{z}_{\rm v}} $.
Thereby, the output current is obtained in terms of the internal voltage as
$    \phavec i_{\rm o} = \left( \mat I_{n} + \phamat{Y}_{\rm c}{\pha{z}_{\rm v}} \right)^{-1} \phamat{Y}_{\rm c} \phavec{\hat{v}}_{\mu} - \left( \mat I_{n} + \phamat{Y}_{\rm c}{\pha{z}_{\rm v}} \right)^{-1} \phavec{y} \pha v_{\rm g}$. Thus, the nonsingular admittance matrix of the augmented network is obtained as $\phamat{\Tilde{Y}}_{\rm c} = \left( \mat I_{n} + \phamat{Y}_{\rm c}{\pha{z}_{\rm v}} \right)^{-1} \phamat{Y}_{\rm c}$. A similar relationship can be obtained between the admittance matrix of the augmented microgrid network $\phamat{\Tilde{Y}}_{\rm m}$ and the original microgrid network $\phamat{{Y}}_{\rm m}$.
We note that fault occurrences and fault clearances in the network can change the topology and parameters of the network. These events can also be taken into account when obtaining the admittance matrix of the Kron-reduced augmented network $\phamat{\Tilde{Y}}_{\rm red}$.

\subsection{Conditions for Transient Stability}

Since the saturation-informed control strategy, under current saturation, results in an equivalent circuit and an equivalent GFM controller that is similar to that of the standard form of the unsaturated GFM control, we can directly extend the stability conditions from previous results in \cite{Xiuqiang_NonlinearStab_dVOC,Xiuqiang_Preprint,he2024passivity}. Thus, transient stability conditions can be obtained for single-converter grid-connected systems, multi-converter grid-connected systems, and also for multi-converter islanded microgrids.

\subsubsection{Single-Converter Grid-Connected Systems} For a single saturation-informed converter grid-connected system, the globally asymptotic stability condition, under current saturation, can be obtained as \cite[Theorem~1]{Xiuqiang_Preprint}
\begin{equation}
\label{condi:globally-stable}
    \Re \bigl\{e^{j\varphi} \tfrac{p^{\star} - jq^{\star}}{{v}^{\star 2}} \bigr\} + \alpha < \tfrac{\alpha}{2} \tfrac{\hat v_{{\mu}\rm s}^2}{v_{\mu}^{\star 2}}  + \Re \bigl\{e^{j\varphi} \pha{y} \bigr\}.
\end{equation}
In \eqref{condi:globally-stable}, we can also replace $v_{\mu}^{\star 2}$ with ${v}^{\star 2}$ since $v_{\mu}^{\star 2} \leq {v}^{\star 2}$. Alternatively, without relying on the information of steady-state voltage $\hat v_{{\mu}\rm s}$, we can directly remove the first term on the right-hand side.

\subsubsection{Multi-Converter Grid-Connected Systems} For a renewable power plant with $n$ saturation-informed GFM converters, the stability condition, under current saturation, can be obtained using the stability condition in \cite[Theorem~1]{he2024passivity},
\begin{equation}
\label{condi:multiGridConnected}
    \max_{k} \Re \bigl\{e^{j\varphi} \tfrac{p_k^{\star} - jq_k^{\star}}{{v}_k^{\star 2}} \bigr\} + \alpha < \tfrac{\alpha}{2}\tfrac{{\hat{v}}_{{\mu}\rm s\min}^2}{v_{\mu\max}^{\star 2}} + \lambda_{\min} \bigl(\Re \bigl\{e^{j\varphi} \phamat{\Tilde{Y}}_{\rm c} \bigr\} \bigr),
\end{equation}
where ${\hat{v}}_{\mu\rm s \min}$ denotes the minimum steady-state voltage of the converters, $v_{\mu\max}^{\star}$ denotes the maximum setpoint, ${\lambda _{\min}}(\cdot)$ denotes the smallest eigenvalue, and $\lambda_{\min} \bigl(\Re \bigl\{e^{j\varphi} \phamat{\Tilde{Y}}_{\rm c} \bigr\} \bigr)$ is known as the generalized short-circuit ratio (gSCR) \cite{Wei_SSSA_PowerElecSys}. The gSCR quantifies the strength of the augmented network, i.e., the network that includes the constant virtual impedance. It can be seen that the results in \eqref{condi:multiGridConnected} are decentralized conditions for parameter tuning and stability guarantees.

\subsubsection{Multi-Converter Islanded Microgrids} For an islanded microgrid with $n$ saturation-informed GFM converters, the decentralized stability condition, under current saturation, can be obtained using the stability condition in \cite[Theorem~2]{Xiuqiang_NonlinearStab_dVOC},
\begin{equation}
\label{condi:multiMicrogrid}
    \max_{k} \Re \bigl\{e^{j\varphi} \tfrac{p_k^{\star} - jq_k^{\star}}{{v}_k^{\star 2}} \bigr\} + \alpha < \tfrac{(1+\cos{\Bar{\delta}})(1-\Bar{\mu})^2}{2} \lambda_2 \bigl(\Re \bigl\{e^{j\varphi} \phamat{\Tilde{Y}}_{\rm m} \bigr\} \bigr)
\end{equation}
where $\bar{\delta} \in [0,\, \pi/2)$ is the maximal steady-state phase difference, $\bar{\mu} \in (0,\, 1)$ is the maximal steady-state voltage ratio deviation from $1$, ${\lambda _2}(\cdot)$ denotes the second smallest eigenvalue, and $\lambda_2 \bigl(\Re \bigl\{e^{j\varphi} \phamat{\Tilde{Y}}_{\rm m} \bigr\} \bigr)$ is known as the algebraic connectivity of the graph corresponding to the real part of the rotated microgrid admittance matrix that is also augmented with the constant virtual impedance.

Under the assumption of uniform network impedance angles and tuning $\varphi$ and $\angle {\pha{z}_{\rm v}}$ to be equal to the impedance angle, the values of the gSCR in \eqref{condi:multiGridConnected} and the algebraic connectivity in \eqref{condi:multiMicrogrid} can be related to their respective original values as: $    \lambda_{\min} (\Re \{e^{j\varphi} \phamat{\Tilde{Y}}_{\rm c} \} ) = {\lambda_{\min}( \mat L_{\rm c} )} ({1 + \lambda_{\min}( \mat L_{\rm c} )\lvert {\pha{z}_{\rm v}} \rvert)^{-1}}$ and $    \lambda_2 (\Re \{e^{j\varphi} \phamat{\Tilde{Y}}_{\rm m} \} ) = {\lambda_2 (\mat L_{\rm m})} {(1 + \lambda_2\bigl( \mat L_{\rm m} \bigr)\lvert {\pha{z}_{\rm v}} \rvert)^{-1}}$, where $ \mat L_{\rm c}  \coloneqq \Re \bigl\{e^{j\varphi} \phamat{{Y}}_{\rm c} \bigr\}$ and $ \mat L_{\rm m} \coloneqq \Re \bigl\{e^{j\varphi} \phamat{{Y}}_{\rm m} \bigr\}$. Here, we observe that a high virtual impedance would have an adverse effect on stability as it reduces the strength of the network.

\subsubsection{Fault Ride-Through and Recovery}
When the fault is cleared, the current saturation should exit and the control should switch/reset back to the original complex-droop control and voltage control. Considering Example~\ref{specialCase}, we show that it is easy to accomplish this switchback even if we only detect $\mu_{\rm f} = 1$ is reached. 

Consider that the set of equations in \eqref{eq:saturation-feedback-control} has a current-saturated solution. When the grid voltage $v_{\rm g}$ recovers, the solution will be desaturating, i.e., $\mu_{\rm s}^2 \geq 1$ if $\lambda_{\rm exsat} \coloneqq \alpha \frac{(v_{\rm g} + i_{\lim}\lvert \pha {z} \rvert)^2}{v^{\star 2}} + \frac{i_{\lim}}{v_{\rm g} + i_{\lim}\lvert \pha {z} \rvert} - \sigma_{\varphi}^{\star} - \alpha \geq 0 $. Following reasoning similar to that used to examine the existence of a current-saturated equilibrium, it is straightforward to show that $\lambda_{\rm exsat} \geq 0$ implies a solution of \eqref{eq:saturation-feedback-control} with $\mu_{\rm s}^2 \geq 1$. However, for more reliable switching during transients, we recommend using the combination of multiple signals.

\subsection{Comparison With Existing Current-Limiting Strategies}

We compare the proposed saturation-informed strategy with two existing typical current-limiting strategies: \emph{threshold virtual impedance} (also called adaptive virtual impedance) and \emph{current limiter with virtual admittance}. The \emph{threshold virtual impedance strategy} employs an explicit virtual impedance emulation module to reduce the voltage reference whenever the current exceeds a threshold $i_{\mathrm{th}}$. Since the module is typically located outside the inner voltage and current control loops, the current-limiting speed is limited by the control bandwidth of the inner loops. Moreover, it requires careful parameter tuning to ensure that the current is limited within $[i_{\mathrm{th}},\, i_{\mathrm{lim}}]$. Compared to the \emph{current limiter with virtual admittance} in Table~\ref{tab:comparison-current-limiting-strategies}, our proposed strategy further incorporates saturation-informed feedback in the virtual admittance control loop. The crucial benefit is that the equivalent virtual impedance is constant and the equivalent internal virtual voltage varies its magnitude adaptively, cf. the equivalent circuit in Fig.~\ref{fig:EqCktdVOCSatCirc}(b). This equivalent circuit allows us to obtain an equivalent grid-forming controlled system, consequently enabling the extension of the existing stability results from current unsaturated conditions to current saturated conditions. Moreover, unlike our proposed strategy, both the pre-existing current-limiting strategies result in current-dependent virtual impedance, which significantly increases the difficulty of transient stability analysis, especially for multi-inverter interconnected systems. Since the resulting virtual impedance is current-dependent, it may become quite large under severe fault, thus destabilizing the system \cite{zhang2023current}. The comparison is summarized in Table~\ref{tab:comparison-current-limiting-strategies}.

\begin{table*}
\centering
\caption{Comparison Between the Proposed Saturation-Informed Strategies With Existing Current-Limiting Strategies.}
\begin{tabular}{llll}
\hline\hline
& \makecell[l]{Threshold/adaptive virtual impedance \\ \cite{paquette2015virtual,qoria2020current}} & 
\makecell[l]{Current limiter with virtual admittance \\ \cite{rosso2021implementation,fan2022equivalent,zhang2023current,saffar2023impacts,laba2023virtual}} & 
\makecell[l]{Proposed saturation-informed strategy \\ This work} \\
\hline
\makecell[l]{Current-limiting \\ strategy design} & \makecell[l]{Increase the virtual impedance whenever \\ the current exceeds a threshold $i_{\mathrm{th}}$} & 
\makecell[l]{Current limiter saturates reference and \\ virtual admittance acts as proportional \\ control for anti-windup} & 
\makecell[l]{Current limiter saturates reference, \\ virtual admittance acts as proportional \\ control for anti-windup, and saturation \\ feedback renders a constant impedance} \\
\arrayrulecolor{black!10}\hline
\makecell[l]{Current-limiting speed} & 
Slow & Fast & Fast \\
\hline
\makecell[l]{Overcurrent utilization} & \makecell[l]{No, limited within $[i_{\mathrm{th}},\, i_{\mathrm{lim}}]$} & Yes, limited at $i_{\mathrm{lim}}$ & Yes, limited at $i_{\mathrm{lim}}$ \\
\hline
\makecell[l]{Tuning complexity} & Complicated & Simple & Simple \\
\hline
\makecell[l]{Resulting impedance} & Current-dependent & Current-dependent & Constant \\
\hline
\makecell[l]{Transient stability \\ analysis} & 
\makecell[l]{Difficult since the resulting virtual \\ impedance is current-dependent } & \makecell[l]{Difficult since the resulting virtual \\ impedance is current-dependent } & 
\makecell[l]{Tractable since the equivalent system \\ allows extending prior analysis methods} \\
\arrayrulecolor{black}
\hline \hline
\end{tabular}
\label{tab:comparison-current-limiting-strategies}
\end{table*}

\section{Simulation Validation}\label{sec:Sim_Results}

In this section, the transient instability of the conventional complex-droop control, under current saturation, is demonstrated through detailed time-domain simulations. In comparison, the saturation-informed complex-droop control, under current saturation, is shown to be stable. We implement the inner controllers in the $dq$ coordinate frame to validate that the analysis is valid for either coordinate. To do so, the voltage reference $\hat{\pha v}$ is transformed from the $\alpha\beta$ to the $dq$ coordinates using the reference angle $\theta = \tan^{-1}({\hat{v}_{\beta}}/{\hat{v}_{\alpha}})$.

\subsection{Case Study I: Single-Converter Grid-Connected Systems}

To validate the transient stability of the proposed saturation-informed control strategy, time-domain simulations are first performed on a single-converter grid-connected system shown in Fig.~\ref{fig:ComplexDroopSimple}. Base values are chosen as $V_{\textrm{base}}=690\,\si{V}$ and $S_{\textrm{base}}=2\,\si{MVA}$. Grid resistance and inductance are set as $r_{\rm g}=0.1\,{\rm p.u.},\,l_{\rm g}=0.1\,\rm{p.u.}$. The converter setpoints are set as $p^{\star}=0.2\,{\rm p.u.},\,q^{\star}=0.4\,\rm{p.u.}$ and ${v}^{\star}=1.0\,\rm{p.u.}$. The current limit is chosen as $i_{\lim}=1.1\,{\rm p.u.}$ The converter parameters are specified as: $\varphi=\pi/4$~rad, $\eta = 0.04 $~p.u., $\alpha = 5 $~p.u., $\tau= 0.1$~s, $k_{\rm p}^{\rm v} = 5$, {$k_{\rm r}^{\rm v} = 10$, {$k_{\rm p}^{\rm c} = 1$, $k_{\rm r}^{\rm c} = 10$. The grid voltage dips from $1.0\,{\rm p.u.}$ to $0.3\,{\rm p.u.}$ at $t=3\,\si{s}$ and recovers to its original value at $t=4\,\si{s}$. This voltage drop results in current saturation of the GFM converter during this period.

\begin{figure}
    \centering
    \includegraphics{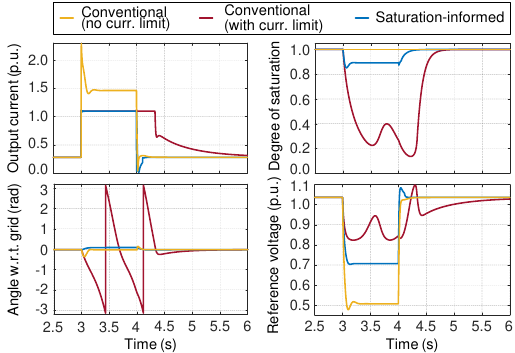}
     \caption{Comparison of the control strategies for a single-converter grid-connected system under a grid fault.}
    \label{fig:SCIB_Results}
\end{figure}

For the saturation-informed converter, the change in the terminal voltage and the DoS signal are used to make the following changes once the current is saturated: 1) The integrator is disabled; 2) the proportional gain $k_{\rm p}^{\rm v} = 5$ is changed to $1/\pha{z}_{\rm v} = 5e^{-j\pi/4}$; 3) the power setpoint is changed to $\phaconj{\varsigma}^{\star} = \frac{p^{\star} - jq^{\star}}{{{v}^{\star 2}}} = 0.2-0.2j$. The parameters follow the tuning described in Example 1 and therefore theoretically ensure that there exists an equilibrium during current saturation. In our simulation, the change in the terminal voltage is used to re-enable the integrator and switch back the parameters to their original values when the converter exits current saturation. 

The simulation results are shown in Fig.~\ref{fig:SCIB_Results}. The conventional complex-droop GFM converter, with the circular current limiter, is unable to maintain transient stability during the period of current saturation and experiences angle drift, i.e., loss of synchronization with the grid. However, the saturation-informed GFM converter is able to maintain transient stability during current saturation, similar to the case of the complex-droop-controlled converter without any current limiter. The proposed control strategy also has good FRT capabilities.

\begin{figure}
    \centering
    \includegraphics{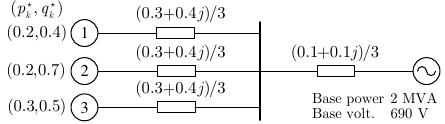}
     \caption{Three-converter grid-connected systems with arbitrary power setpoints.}
    \label{fig:3CIB_model}
\end{figure}

\begin{figure}
    \centering
    \includegraphics{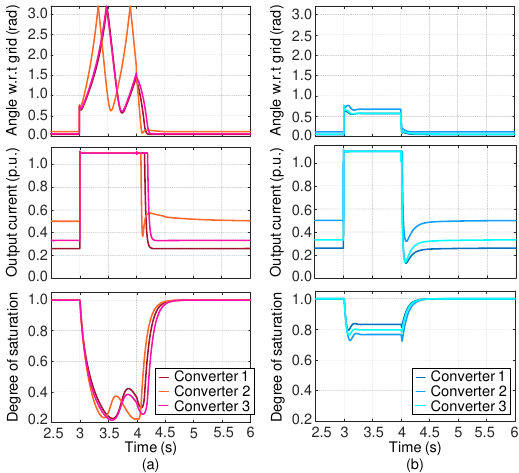}
     \caption{Comparison of (a) the conventional GFM control and (b) the saturation-informed control for the multi-converter system under a grid fault.}
    \label{fig:MCIB_Results}
\end{figure}

\subsection{Case Study II: Multi-Converter Grid-Connected Systems}

In order to validate the transient stability of the proposed control strategy in \emph{multi-converter} grid-connected systems, simulations are performed on a test system with three converters, as shown in Fig.~\ref{fig:3CIB_model}. The base values of the system and the converter parameters are the same as in the single-converter grid-connected example. However, we use arbitrarily specified power setpoints, unit voltage setpoints, and real-valued $\frac{1}{\pha{z}_{\rm v}}$ to validate the robustness of the proposed current-liming control. The grid voltage dips from $1.0\,{\rm p.u.}$ to $0.1\,{\rm p.u.}$ at $t=3\, \si{s}$ and recovers to its original value at $t=4\, \si{s}$. This voltage dip results in the current saturation of all the converters during this period.

For the saturation-informed GFM converters, the change in the terminal voltage is used to disable or re-enable the integrator when the current is saturated or unsaturated, respectively. The other parameters of the converters remain unchanged. The saturation-informed current-limiting control for GFM converters is able to maintain transient stability under current saturation, unlike the conventional complex-droop GFM control strategy, as seen in the comparative simulation results in Fig.~\ref{fig:MCIB_Results}.

\begin{figure}
    \centering
    \includegraphics{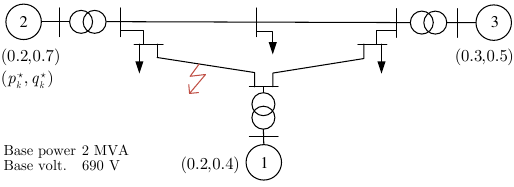}
     \caption{Single-line diagram of the IEEE 9-bus system with GFM converters.}
    \label{fig:9Bus_Model}
\end{figure}

\begin{figure}
    \centering
    \includegraphics{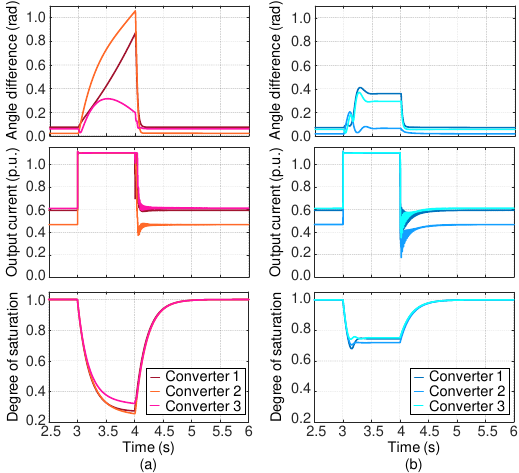}
     \caption{Comparison of (a) the conventional GFM control and (b) the saturation-informed control for the IEEE 9-bus system under a grid fault.}
    \label{fig:9Bus_Results}
\end{figure}

\subsection{Case Study III: IEEE 9-Bus Test System}

Apart from grid-connected scenarios, it is also essential that transient stability in a multi-converter islanded system is maintained under current saturation. To investigate the efficacy of the proposed control strategy in such a system, simulations are performed on the standard IEEE 9-bus test case, as shown in Fig.~\ref{fig:9Bus_Model}. The synchronous machines are replaced with GFM converters of the same rating, and the converter parameters are the same as in the single-converter case. The voltage setpoints are set to unity, and the power setpoints are indicated in Fig.~\ref{fig:9Bus_Model}.

A three-phase fault occurs at $t=3\,\si{s}$ between nodes 4 and 5, and is cleared at $t=4\,\si{s}$. The fault causes severe voltage dips, and therefore the saturation of current for the three converters. From Fig.~\ref{fig:9Bus_Results}, it can be seen that the proposed saturation-informed GFM control renders the system stable, while with conventional complex-droop GFM converters, transient instability is experienced during the grid fault.

\section{Conclusion}\label{sec:Conc}

This paper investigates the transient stability of complex-droop-controlled GFM converters when circular current limiters are used for current limiting. A novel saturation-informed complex-droop control strategy is presented to ensure transient stability under current saturation, which leverages the information of the level of saturation. Parametric stability conditions are obtained for the proposed control strategy and guidelines for parameter tuning are also provided. The efficacy of the control strategy to limit current while maintaining transient stability under grid faults is validated by high-fidelity EMT simulations on various test cases. Our future work will investigate the current-limiting performance and transient stability of the developed strategy on other types of grid-forming controls and the extension to asymmetrical grid fault conditions.

\bibliographystyle{IEEEtran}
\bibliography{IEEEabrv,Bibliography}

\end{document}